\documentstyle[12pt,psfig,epsfig,supertabular]{article}
\newcommand\numu{{\nu_\mu}}
\newcommand\anumu{\bar\nu_\mu}

\newcommand{\beq}{\begin{equation}}
\newcommand{\eeq}{\end{equation}}
\newcommand{\bea}{\begin{eqnarray}}
\newcommand{\eea}{\end{eqnarray}}
\def\dm2{\Delta m^2}
\def\sq2{sin^2(2\Theta)}
\def\nubar{\overline {\nu} }

\newcommand{\FLUKA}{{\sc FLUKA}}

\begin{document}     

\title{\bf High energy extension of the \FLUKA~atmospheric neutrino flux }

\author{G.~Battistoni$^1$, A.~Ferrari$^2$, T~Montaruli$^3$ and P.R.~Sala$^4$}

\maketitle

\begin{center}
{\it 
(1) INFN, Sezione di Milano, 20133 Milano, Italy\\
(2) CERN, Geneva 23, Switzerland, on leave of absence from INFN Milano \\
(3) University of Bari and INFN, 70126 Bari, Italy\\
(4) ETH Zurich, Switzerland, on leave of absence from INFN Milano\\
}
\end{center}

\begin{abstract}
The atmospheric neutrino flux calculated with
\FLUKA{} was originally limited to 100$\div$200 GeV for statistical reasons.
In order to make it available for the analysis of high energy events, 
like upward through-going muons detected by $\nu$ telescopes, 
we have extended the calculation so to provide a reliable neutrino yield
per primary nucleon up to about 10$^6$ GeV/nucleon, as far as the
interaction model is concerned. We point out that the primary flux model
above 100 GeV/nucleon still contributes with an important
systematic error to the $\nu$ flux.
\end{abstract}

\section{Introduction}
The main motivation of
\FLUKA{}\cite{fluka} calculation of atmospheric neutrino flux\cite{flukanu}
is the attempt to minimize as much as possible the theoretical uncertainties 
connected to the shower and hadronic interaction models.
The high accuracy of the algorithms adopted in \FLUKA{} and the wide
set of experimental data used to benchmark its hadronic and
electro-magnetic interaction models have already allowed
the use of this simulation models in many applications where the
maximum available precision
is requested.
\FLUKA{} has been successfully used also in cosmic ray physics. 
For this reason it is also going to be adopted as an option
in the CORSIKA code, at least for interactions below 80 GeV\cite{heck}.
Our results concerning 
the atmospheric neutrino flux calculations have been presented
in\cite{flukanu} and references therein . There, the other important factors 
contributing to the systematic uncertainties which are not connected to
the \FLUKA{} code, are
the atmosphere description, the accuracy of the geomagnetic model, and, as the
most important contribute, the knowledge of the primary flux.

The practical problems arising from our approach is that the demand of
computer power for very high energy showers may be very large. This is the
reason why so far we have presented calculation results only up to about
100 GeV (useful for the analysis
of contained events) due to insufficient statistics at higher energy. 
The low energy sector was also considered
less affected by the uncertainty in the primary flux thanks to the
high accuracy data from AMS\cite{ams} and BESS\cite{bess}.
This statistical limitation in energy has however prevented the use of 
\FLUKA{} results for a complete analysis of atmospheric neutrinos, including
through--going muons produced by neutrinos with mean energies 50--100
GeV. 
It has been already
shown how this requires the knowledge of neutrino flux up to at least 10$^4$
GeV. In the high energy sector the existing available calculations  
are those coming mostly from faster approaches, like those of
Bartol\cite{bartol} and 
HKKM\cite{hkkm}. Also the new recent HKKM calculation\cite{hkkm2001}
(HKKM2001) stops in 
practice at about 1 TeV and makes use of extrapolations and/or
simplifications.

In order to overcome this limitations, we have performed an extensive 
simulation run so to extend the \FLUKA{}
calculation of atmospheric neutrino fluxes up to 10$^4$ GeV. 
This has been done in such a way to consider different possible primary
spectra. Results are presented in this paper.

\section{The Calculation Model}
We have used the same calculation set-up reported in \cite{flukanu}, where
also some 
details about the physical model are reported. Here the different
strategy is to generate neutrinos avoiding the sampling from a unique
primary all-nucleon spectrum and instead prepare parallel simulation runs 
for many
different small energy intervals, summing together the results with a
proper weighting scheme so to reproduce any desired input spectrum. In this
way we could obtain a statistical uncertainty almost independent on energy.
We started from few tens of GeV of primary energy and 
in order to achieve reliable results at a maximum neutrino energy of 10$^4$
GeV, we have pushed the maximum nucleon energy up to 10$^6$ GeV.  
The relatively high energy allowed some simplification, avoiding the
question of solar modulation and geomagnetic cutoff. 
Furthermore, the angular distribution for non oscillated flux has been
recorded in 10 bins in a single hemisphere.

\section{Results and Discussion}
We have chosen two different primary all-nucleon spectra
to present and compare the calculation results. 
The first one is
the 2001 Bartol fit of ref.~\cite{bart2001} taking into account the most
recent 
direct measurements, and that was used as reference in our last
work\cite{flukanu}.
The second one is the previous
Bartol fit to all-nucleon flux of ref.~\cite{bartol}, which was already
used by 
us for the first (before 2001) \FLUKA{} calculations.
The updated flux tables, obtained using the 2001 Bartol fit,
are available from our web site~\cite{nuweb}. Numerical values of the
$\numu$
and $\anumu$ flux
above 20 GeV are reported
in Tables \ref{tab1} and \ref{tab2}. In Tables\ref{tab3} and \ref{tab4} we
instead 
report the neutrino flux obtained using the old Bartol fit to primary flux 
of ref.\cite{bartol}.

\begin{figure}[p]
  \begin{center}
    \includegraphics[height=15cm]{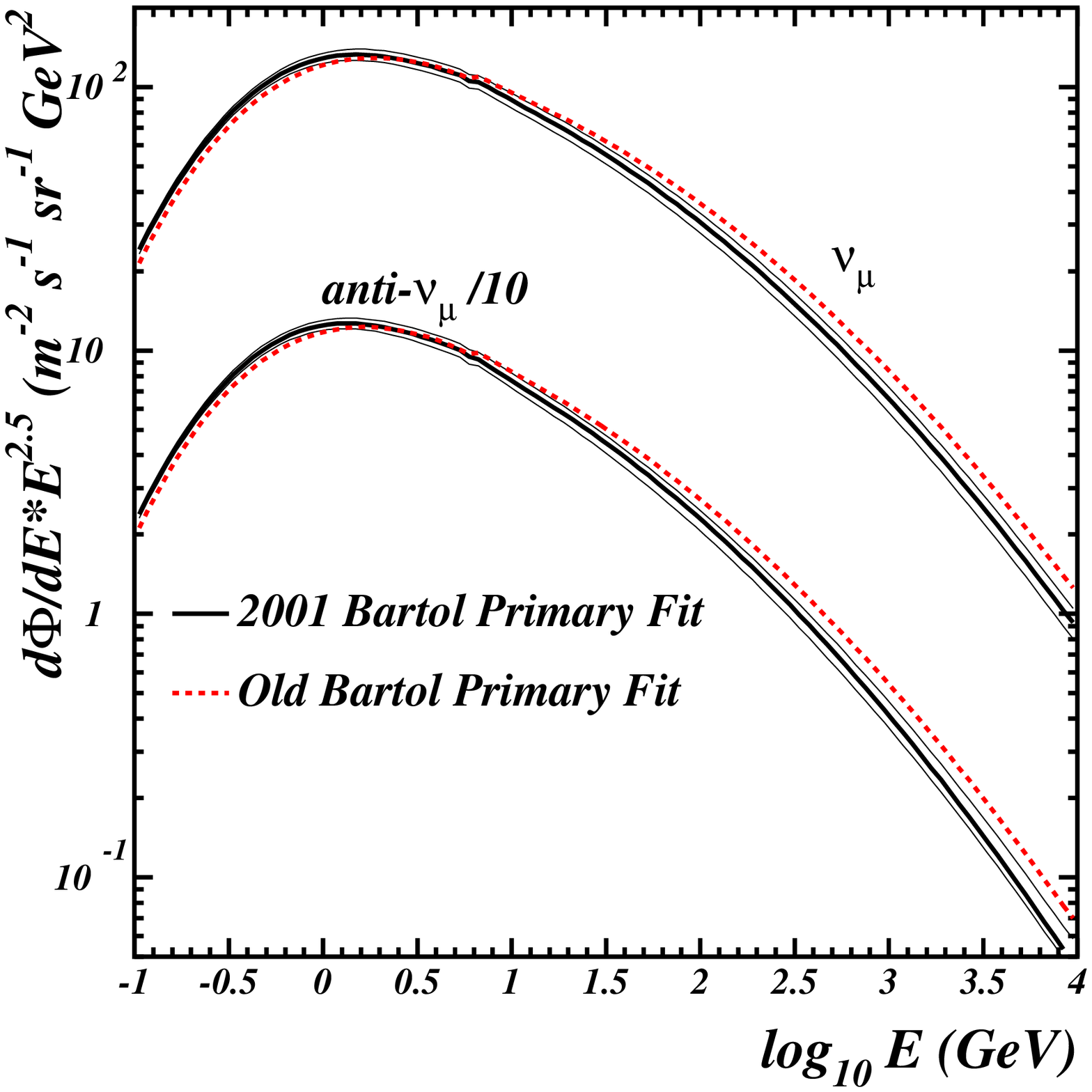} %%%%% ``13.5pc'' is
  \caption{Angle integrated (and without oscillations)
    atmospheric $\numu$ and $\anumu$ \FLUKA{} fluxes for the 2
    choices of primary spectrum considered here, weighted by
    E$^{2.5}$.\label{fig1}}
  \end{center}
\end{figure}

In
Fig.\ref{fig1} we show the angle integrated flux for muon $\nu$ and
$\nubar$ for the two primary spectra. 
It is evident that in the high energy sector the choice between the two
different options for the primary all-nucleon spectrum has a large impact.

\begin{figure}
  \begin{center}
  \begin{tabular}{cc}
    \includegraphics[width=7cm]{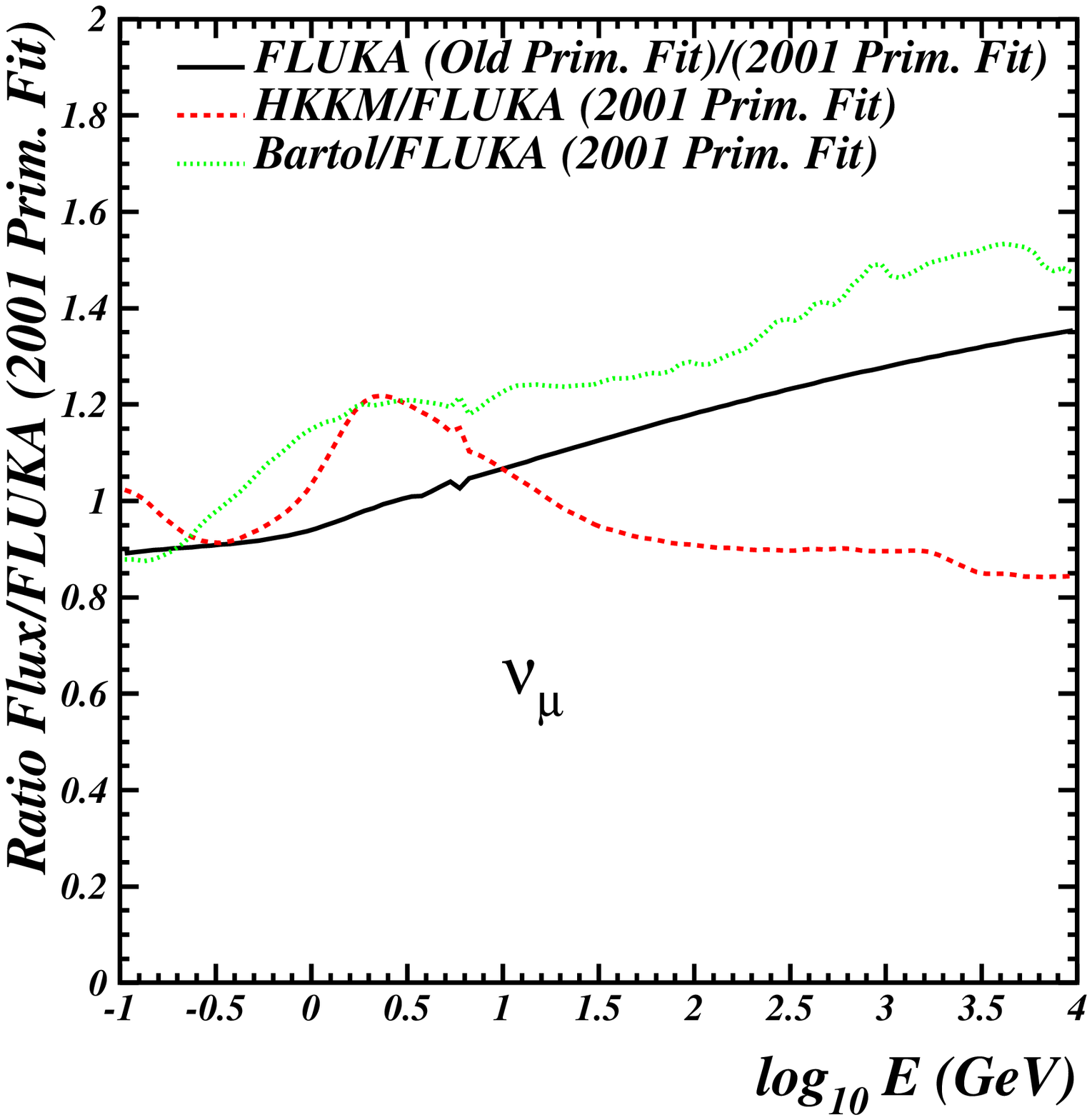} &
    \includegraphics[width=7cm]{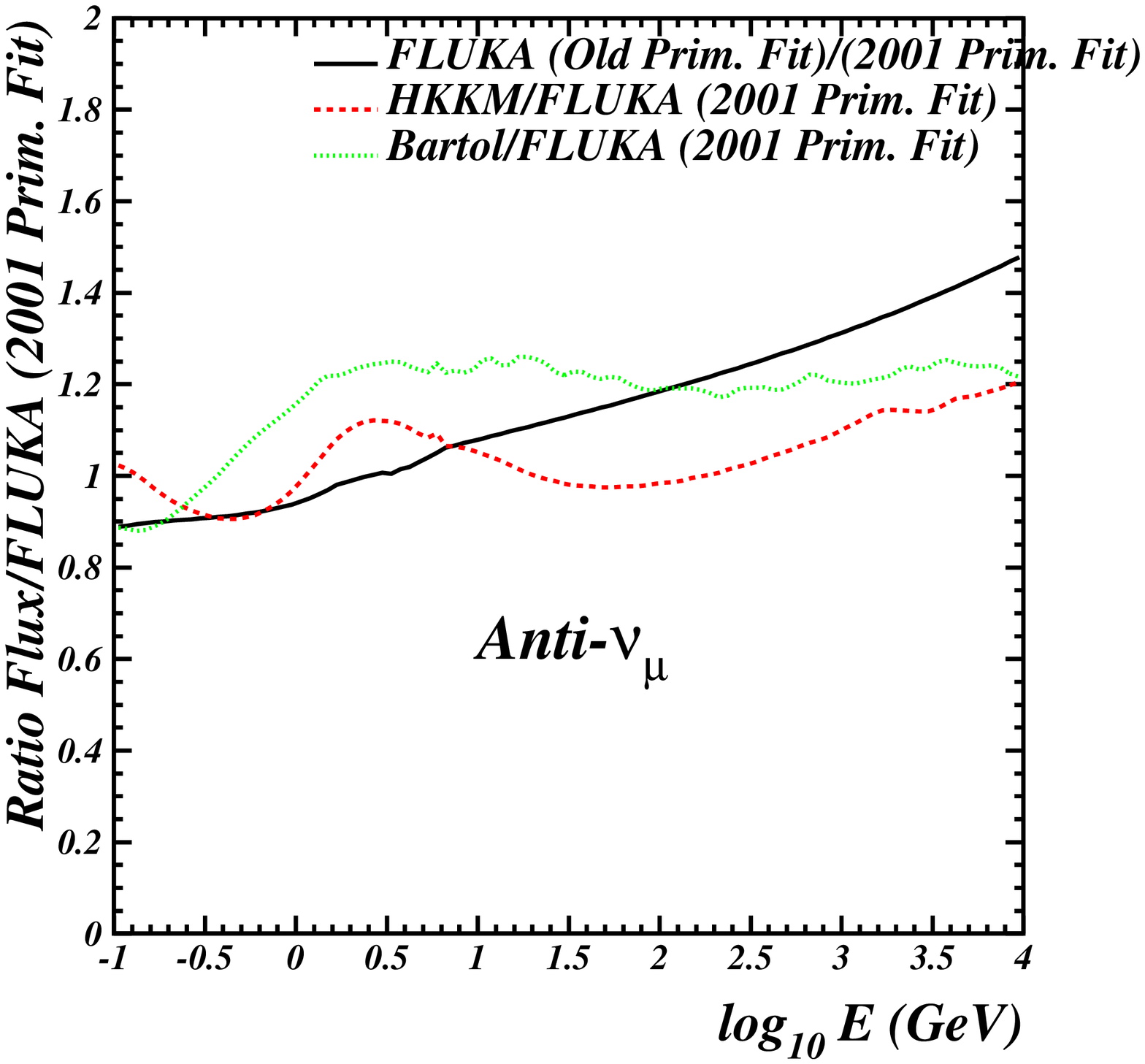} \\
  \end{tabular}
  \vspace{-0.5pc}
  \caption{{\it Left}: $\numu$ flux ratio vs. energy with respect to
    the FLUKA + 2001 primary fit of: 1) the FLUKA+Old Bartol primary fit
    (solid line), Bartol flux (dotted green line) and HKKM2001 
    flux (dashed red line). 
    {\it Right}: the same for $\anumu$.\label{fig2}}
  \end{center}
\end{figure}

In order to give a quantitative figure of this statement in Fig.\ref{fig2}
we also show, as a function of neutrino energy, 
the ratio of the flux calculated
using the old Bartol fit with
respect to the one obtained with the 2001 fit (solid line) for $\numu$
and $\anumu$. In the same
plots we also show the ratio of the Bartol and HKKM2001 flux with respect
to FLUKA with the 2001 primary fit. 

\begin{figure}[ht]
  \begin{center}
  \begin{tabular}{cc}
    \includegraphics[width=7cm]{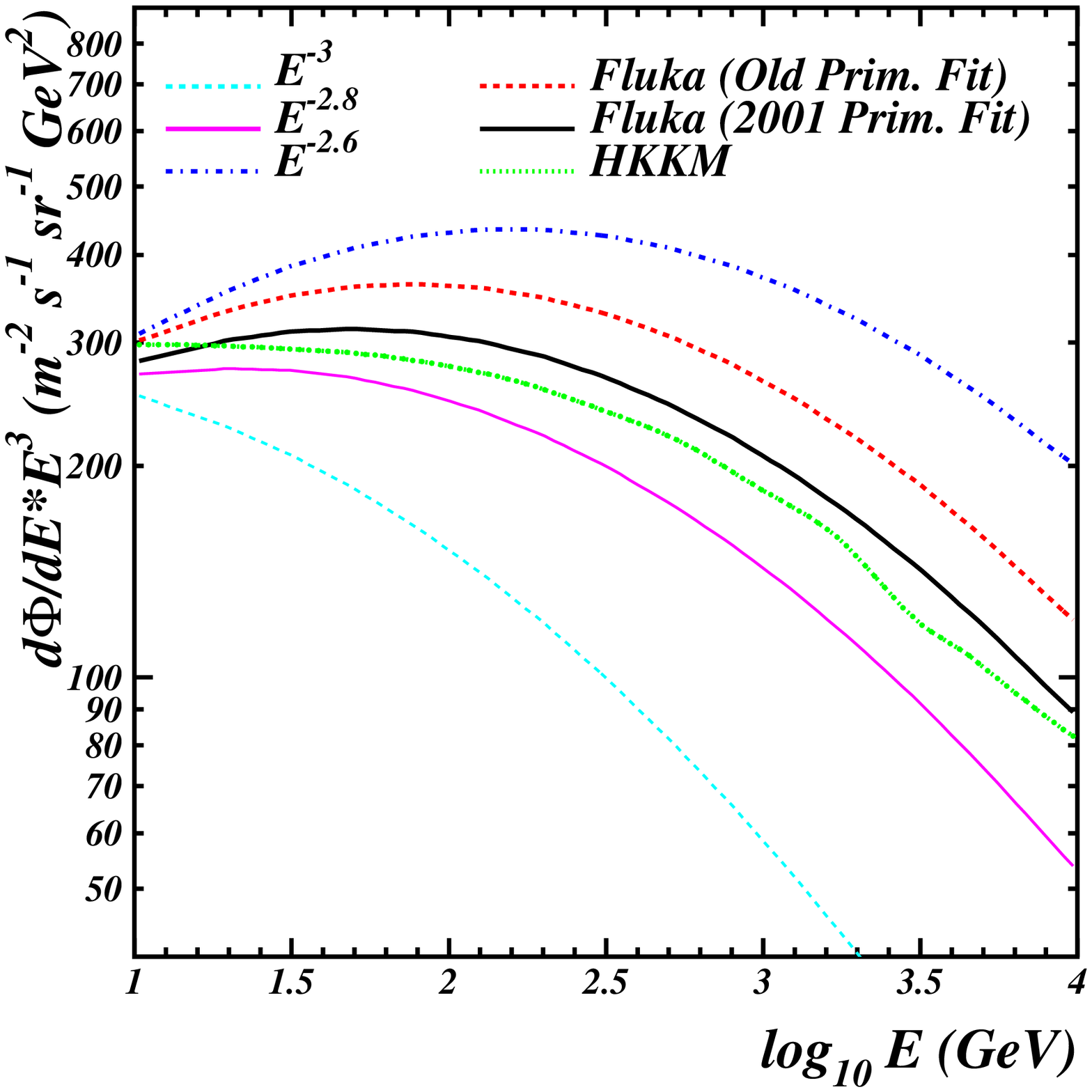} & 
    \includegraphics[width=7cm]{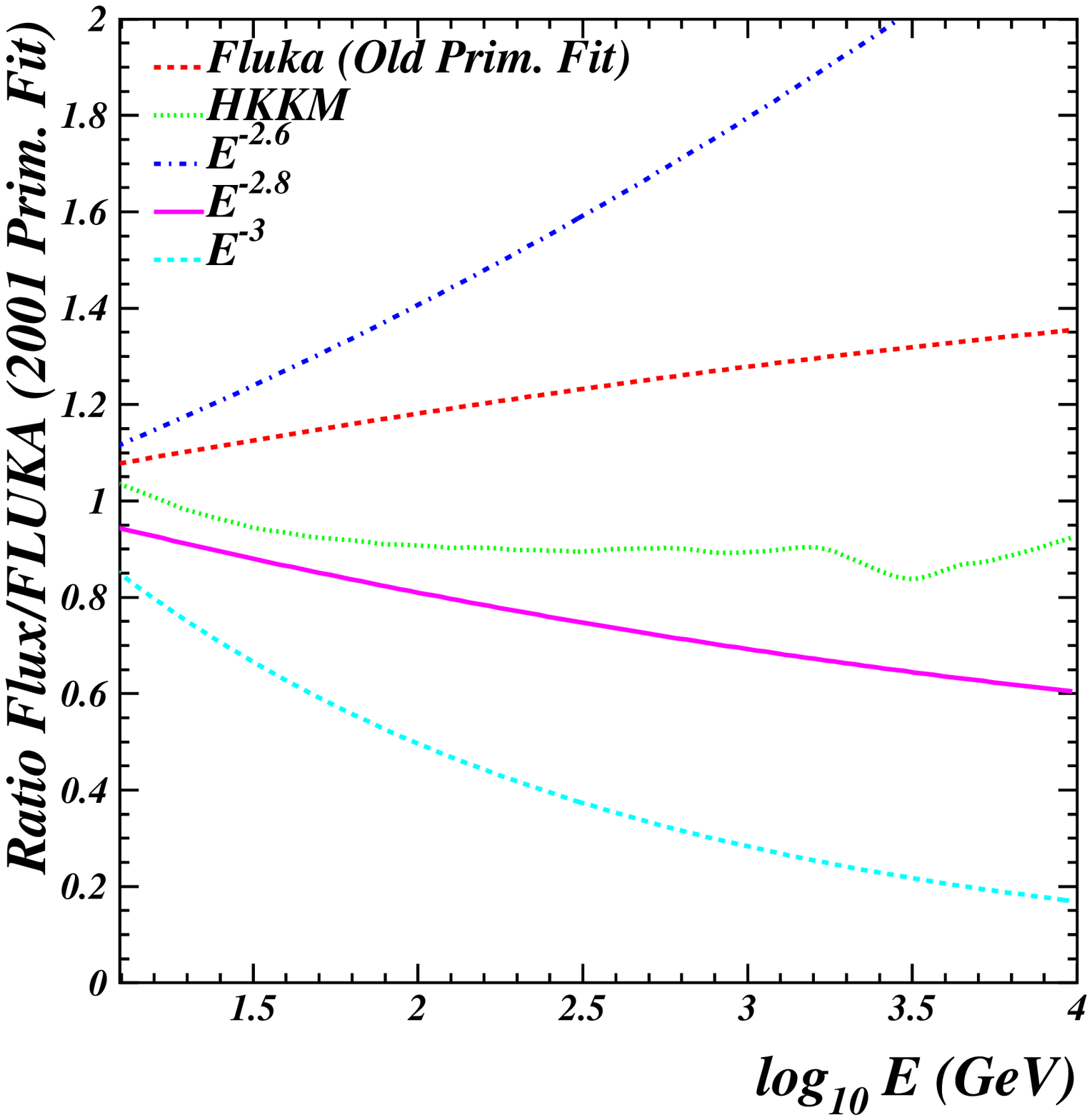} \\ 
  \end{tabular}
  \vspace{-0.5pc}
  \caption{{\it Left}: Angle averaged $\numu$ \FLUKA{} fluxes weighted by E$^3$,
    including also the case of
    simple power law primary spectra with different spectral indexes,
    ranging from 2.6 (upper curve) to 3.0 (lower curve). The
    HKKM2001 flux is also shown. {\it Right}: ratios of the
    different cases with respect to the present \FLUKA{} result.
    \label{fig3}}
  \end{center}
\end{figure}

The difference introduced by using the two primary spectra
evidently results from a different weight given to the
high energy primary measurements. The highest energy point in the
recent high precision experimental measurements from AMS and BESS is 
at 100 GeV. Higher
energy points in the TeV region (essentially from JACEE\cite{jacee} and
RUNJOB\cite{runjob}) 
have much larger error bars. There are indeed suspicions that the
multi--TeV proton component could be underestimated in the 2001 fit.
Assuming to consider the 100 GeV point as a pivot from which a simple power
law spectrum ($E^{-\gamma}$) is used to extrapolate to the high energy region,
we obtain the results
shown in Fig.\ref{fig3}, by varying the spectral index $\gamma$ from 2.6
to 3.0. We also show the flux ratios.

\section{Conclusions}
The \FLUKA{} atmospheric neutrino spectrum calculation has been extended up
to 10$^4$ GeV, so that it can be used for the analysis of upward going muon
experiments. Above 1 TeV, the uncertainty in the primary flux seems 
to be the most important contribution: it may exceed 20\%.
In principle this has no impact on the essential feature of experimental 
analysis, namely the deformation of the
angular distribution of upward going muons as induced by oscillations.
However, also the absolute normalization has some importance, at least to
achieve a better internal consistency. As a first guess, 
considering the event rate of
upward--going muons observed by MACRO\cite{macro} and
Super--Kamiokande\cite{superk}, 
the results that we obtain using the 2001 Bartol fit seems to exhibit a too
low normalization (of the order of 15$\div$20 \%, if we include
oscillations), although this has to be confirmed 
by detailed analysis performed by the experimental collaborations.
Within the statistical uncertainty of the 2001 primary fit, our results are
very close to those of the HKKM2001 calculation. A less steep primary
spectrum at high energy, as that of ref.~\cite{bartol}, would probably
provide a better absolute normalization.

\section*{Acknowledgments}
We are grateful to the Bartol group (T.K. Gaisser et al.) for providing us
with the
fits to primary spectrum and to M. Honda for the HKKM flux tables
and related information.

\begin{table}[ht]
\begin{center}
\begin{tabular}{|r||c|c|c|c|c|c|}
\hline 
\multicolumn{1}{|c||}{$\phi (\numu)$}  
& \multicolumn{6}{|c|}{ Cosine of zenith angle} \\
\hline
 E$_\nu$ (GeV)  & -0.975 & -0.825 & -0.625 & -0.425 & -0.225 & -0.025  \\
\hline 
\hline
   26.607&0.13E-05&0.13E-05&0.14E-05&0.16E-05&0.19E-05&0.25E-05 \\
   37.584&0.44E-06&0.46E-06&0.51E-06&0.57E-06&0.69E-06&0.91E-06 \\
   53.088&0.15E-06&0.16E-06&0.18E-06&0.20E-06&0.25E-06&0.33E-06 \\
   74.989&0.53E-07&0.56E-07&0.63E-07&0.70E-07&0.88E-07&0.12E-06 \\
  105.925&0.18E-07&0.19E-07&0.22E-07&0.24E-07&0.31E-07&0.43E-07 \\
  149.624&0.60E-08&0.64E-08&0.74E-08&0.83E-08&0.11E-07&0.15E-07 \\
  211.349&0.20E-08&0.21E-08&0.25E-08&0.28E-08&0.37E-08&0.53E-08 \\
  298.538&0.66E-09&0.70E-09&0.83E-09&0.94E-09&0.13E-08&0.18E-08 \\
  421.697&0.21E-09&0.23E-09&0.27E-09&0.31E-09&0.42E-09&0.62E-09 \\
  595.662&0.67E-10&0.73E-10&0.88E-10&0.10E-09&0.14E-09&0.21E-09 \\
  841.395&0.21E-10&0.23E-10&0.28E-10&0.33E-10&0.47E-10&0.70E-10 \\
 1188.503&0.66E-11&0.72E-11&0.89E-11&0.11E-10&0.15E-10&0.23E-10 \\
 1678.805&0.20E-11&0.22E-11&0.28E-11&0.34E-11&0.50E-11&0.74E-11 \\
 2371.374&0.62E-12&0.68E-12&0.86E-12&0.11E-11&0.16E-11&0.24E-11 \\
 3349.656&0.19E-12&0.20E-12&0.26E-12&0.34E-12&0.51E-12&0.76E-12 \\
 4731.514&0.55E-13&0.61E-13&0.79E-13&0.10E-12&0.16E-12&0.24E-12 \\
 6683.441&0.16E-13&0.18E-13&0.23E-13&0.32E-13&0.51E-13&0.74E-13 \\
 9440.610&0.47E-14&0.52E-14&0.69E-14&0.97E-14&0.16E-13&0.23E-13 \\
\hline
\hline
\end{tabular}
\caption{Muon neutrino flux in units of (cm$^{2}$ s sr GeV)$^{-1}$ above 
20 GeV  as a function of energy for different values
of the cosine of zenith angle, for the case of the 2001 Bartol fit of
primaries\protect\cite{bart2001}.\label{tab1}} 
\end{center}
\end{table}

\begin{table}[ht]
\begin{center}
\begin{tabular}{|r||c|c|c|c|c|c|}
\hline 
\multicolumn{1}{|c||}{$\phi (\anumu)$} & \multicolumn{6}{|c|}{ Cosine of zenith angle} \\
\hline
 E$_{\overline {\nu} }$ (GeV)  & -0.975 & -0.825 & -0.625 & -0.425 & -0.225 & -0.025  \\
\hline 
\hline
   26.607&0.94E-06&0.99E-06&0.11E-05&0.13E-05&0.16E-05&0.22E-05 \\
   37.584&0.32E-06&0.34E-06&0.38E-06&0.45E-06&0.57E-06&0.81E-06 \\
   53.088&0.11E-06&0.12E-06&0.13E-06&0.15E-06&0.20E-06&0.30E-06 \\
   74.989&0.37E-07&0.40E-07&0.45E-07&0.53E-07&0.69E-07&0.11E-06 \\
  105.925&0.12E-07&0.13E-07&0.15E-07&0.18E-07&0.24E-07&0.37E-07 \\
  149.624&0.40E-08&0.44E-08&0.50E-08&0.59E-08&0.80E-08&0.13E-07 \\
  211.349&0.13E-08&0.14E-08&0.16E-08&0.20E-08&0.27E-08&0.43E-08 \\
  298.538&0.42E-09&0.46E-09&0.53E-09&0.64E-09&0.90E-09&0.14E-08 \\
  421.697&0.13E-09&0.15E-09&0.17E-09&0.21E-09&0.30E-09&0.47E-09 \\
  595.662&0.42E-10&0.46E-10&0.54E-10&0.65E-10&0.97E-10&0.15E-09 \\
  841.395&0.13E-10&0.14E-10&0.17E-10&0.21E-10&0.31E-10&0.48E-10 \\
 1188.503&0.40E-11&0.43E-11&0.53E-11&0.64E-11&0.10E-10&0.15E-10 \\
 1678.805&0.12E-11&0.13E-11&0.16E-11&0.20E-11&0.32E-11&0.46E-11 \\
 2371.374&0.37E-12&0.39E-12&0.50E-12&0.60E-12&0.10E-11&0.14E-11 \\
 3349.656&0.11E-12&0.11E-12&0.15E-12&0.18E-12&0.31E-12&0.41E-12 \\
 4731.514&0.33E-13&0.33E-13&0.45E-13&0.54E-13&0.96E-13&0.12E-12 \\
 6683.441&0.95E-14&0.94E-14&0.13E-13&0.16E-13&0.29E-13&0.34E-13 \\
 9440.610&0.28E-14&0.27E-14&0.38E-14&0.46E-14&0.89E-14&0.96E-14 \\
\hline
\hline
\end{tabular}
\caption{Muon anti-neutrino flux in units of (cm$^{2}$ s sr GeV)$^{-1}$
  above 20 GeV  
  as a function of energy for different values of the cosine of zenith
  angle, for the case of the 2001 Bartol fit of
  primaries\protect\cite{bart2001}.\label{tab2}} 
\end{center}
\end{table}

\begin{table}[ht]
\begin{center}
\begin{tabular}{|r||c|c|c|c|c|c|}
\hline 
\multicolumn{1}{|c||}{$\phi (\numu)$}  
& \multicolumn{6}{|c|}{ Cosine of zenith angle} \\
\hline
 E$_\nu$ (GeV)  & -0.975 & -0.825 & -0.625 & -0.425 & -0.225 & -0.025  \\
\hline 
\hline
   26.607&0.14E-05&0.15E-05&0.16E-05&0.18E-05&0.21E-05&0.27E-05 \\
   37.584&0.50E-06&0.53E-06&0.58E-06&0.65E-06&0.78E-06&0.10E-05 \\
   53.088&0.18E-06&0.19E-06&0.21E-06&0.23E-06&0.28E-06&0.38E-06 \\
   74.989&0.62E-07&0.66E-07&0.73E-07&0.83E-07&0.10E-06&0.14E-06 \\
  105.925&0.21E-07&0.23E-07&0.26E-07&0.29E-07&0.36E-07&0.50E-07 \\
  149.624&0.73E-08&0.78E-08&0.88E-08&0.10E-07&0.13E-07&0.18E-07 \\
  211.349&0.25E-08&0.26E-08&0.30E-08&0.35E-08&0.45E-08&0.63E-08 \\
  298.538&0.81E-09&0.88E-09&0.10E-08&0.12E-08&0.15E-08&0.22E-08 \\
  421.697&0.27E-09&0.29E-09&0.34E-09&0.40E-09&0.52E-09&0.75E-09 \\
  595.662&0.86E-10&0.93E-10&0.11E-09&0.13E-09&0.18E-09&0.25E-09 \\
  841.395&0.27E-10&0.30E-10&0.36E-10&0.44E-10&0.59E-10&0.85E-10 \\
 1188.503&0.86E-11&0.93E-11&0.11E-10&0.14E-10&0.20E-10&0.28E-10 \\
 1678.805&0.27E-11&0.29E-11&0.36E-11&0.46E-11&0.64E-11&0.91E-11 \\
 2371.374&0.82E-12&0.89E-12&0.11E-11&0.15E-11&0.21E-11&0.29E-11 \\
 3349.656&0.25E-12&0.27E-12&0.34E-12&0.47E-12&0.67E-12&0.93E-12 \\
 4731.514&0.74E-13&0.80E-13&0.10E-12&0.15E-12&0.21E-12&0.29E-12 \\
 6683.441&0.22E-13&0.23E-13&0.31E-13&0.45E-13&0.67E-13&0.91E-13 \\
 9440.610&0.64E-14&0.68E-14&0.93E-14&0.14E-13&0.21E-13&0.28E-13 \\
\hline
\hline
\end{tabular}
\caption{Muon neutrino flux in units of (cm$^{2}$ s sr GeV)$^{-1}$ above 
20 GeV  as a function of energy for different values
of the cosine of zenith angle, for the case of the old Bartol fit of
primaries\protect\cite{bartol}.\label{tab3}} 
\end{center}
\end{table}

\begin{table}[ht]
\begin{center}
\begin{tabular}{|r||c|c|c|c|c|c|}
\hline 
\multicolumn{1}{|c||}{$\phi (\anumu)$} & \multicolumn{6}{|c|}{ Cosine of zenith angle} \\
\hline
 E$_{\overline {\nu} }$ (GeV)  & -0.975 & -0.825 & -0.625 & -0.425 & -0.225 & -0.025  \\
\hline 
\hline
   26.607&0.11E-05&0.11E-05&0.12E-05&0.14E-05&0.18E-05&0.24E-05 \\
   37.584&0.37E-06&0.39E-06&0.44E-06&0.51E-06&0.65E-06&0.90E-06 \\
   53.088&0.13E-06&0.14E-06&0.15E-06&0.18E-06&0.23E-06&0.33E-06 \\
   74.989&0.44E-07&0.47E-07&0.53E-07&0.62E-07&0.81E-07&0.12E-06 \\
  105.925&0.15E-07&0.16E-07&0.18E-07&0.21E-07&0.28E-07&0.42E-07 \\
  149.624&0.50E-08&0.54E-08&0.61E-08&0.72E-08&0.97E-08&0.15E-07 \\
  211.349&0.17E-08&0.18E-08&0.20E-08&0.24E-08&0.33E-08&0.50E-08 \\
  298.538&0.54E-09&0.58E-09&0.67E-09&0.81E-09&0.11E-08&0.17E-08 \\
  421.697&0.17E-09&0.19E-09&0.22E-09&0.26E-09&0.37E-09&0.56E-09 \\
  595.662&0.55E-10&0.59E-10&0.70E-10&0.86E-10&0.12E-09&0.18E-09 \\
  841.395&0.17E-10&0.18E-10&0.22E-10&0.27E-10&0.41E-10&0.59E-10 \\
 1188.503&0.54E-11&0.57E-11&0.70E-11&0.87E-11&0.13E-10&0.19E-10 \\
 1678.805&0.17E-11&0.17E-11&0.22E-11&0.27E-11&0.42E-11&0.58E-11 \\
 2371.374&0.51E-12&0.52E-12&0.66E-12&0.85E-12&0.14E-11&0.18E-11 \\
 3349.656&0.15E-12&0.16E-12&0.20E-12&0.26E-12&0.43E-12&0.54E-12 \\
 4731.514&0.46E-13&0.46E-13&0.60E-13&0.79E-13&0.13E-12&0.16E-12 \\
 6683.441&0.13E-13&0.13E-13&0.18E-13&0.24E-13&0.41E-13&0.47E-13 \\
 9440.610&0.39E-14&0.38E-14&0.52E-14&0.70E-14&0.13E-13&0.14E-13 \\
\hline
\hline
\end{tabular}
\caption{Muon anti-neutrino flux in units of (cm$^{2}$ s sr GeV)$^{-1}$
  above 20 GeV  
  as a function of energy for different values of the cosine of zenith
  angle, for the case of the old Bartol fit of
primaries\protect\cite{bartol}.\label{tab4}}
\end{center}
\end{table}

\end{document}